\documentclass[11pt]{article}
\usepackage{geometry}             
\usepackage{graphicx}
\usepackage{amsmath}
\usepackage[colorlinks]{hyperref}
\usepackage{array}
\usepackage{alltt}
\usepackage{amssymb}
\usepackage{epstopdf}

\graphicspath{{figs/}}  

\usepackage{bm}

\title{Convergence of Taylor Transfer Map for Duffing Equation}
\author{D. Kaltchev$^{a,}$\footnote{Corresponding author.}${\;}^{,}$\thanks{TRIUMF is supported via a contribution through the National Research Council Canada.}
\;and\; 
A.~J.~Dragt$^{b,}$\thanks{Work supported in part by U.S. Department of Energy Grant DE-FG02-96ER40949.}
 \\ \\
{\it $^a$\,TRIUMF, 4004 Wesbrook Mall, Vancouver, B.C., Canada V6T 2A3 }\\
\texttt{kaltchev@triumf.ca}\\ \\
{\it $^b$\,Physics Department, University of Maryland, College Park, Maryland 20742, USA }\\ 
\texttt{dragt@umd.edu}
}

\begin{document}
\maketitle
\section*{Abstract}
According to a theorem of Poincar\'{e}, the solutions to differential equations are analytic functions of (and therefore have Taylor expansions in) the initial  conditions and various parameters provided that the right sides of the differential equations are analytic in the variables, the time, and the parameters.  These Taylor expansions, which provide a transfer map $\cal M$ between initial and final conditions, may be obtained, to any desired order, by integration of the {\em complete} variational equations.  As an example of this approach, the convergence of such an expansion is investigated for the Duffing equation stroboscopic map in the vicinity of a infinite period doubling cascade and resulting strange attractor. 

\section{Introduction}
\setcounter{equation}{0}

Consider any set of $m$ first-order differential equations of the form 
\begin{equation}
  \dot{z}_a = f_a(z_1,\cdots,z_m),\quad a= 1,\cdots,m.
\end{equation}
Here $t$ is the independent variable and the quantities $z_1,\cdots,z_m$ are dependent variables.  For notational convenience let us introduce the vector $z\in {\mathbb{R}}^m$ with components $z_1,\cdots,z_m$ and the vector $f\in {\mathbb{R}}^m$ with components $f_1,\cdots,f_m$.  With this notation, the set of equations (1.1) can be written more compactly in the form
\begin{equation}
\dot{z}= f(z,t).
\end{equation}

Let $z^0$ be a vector of initial conditions specified at some initial time $t=t^0$,
\begin{equation}
z(t^0)=z^0.
\end{equation}
Then, under mild conditions imposed on the functions $f_a$ that appear on the right side of (1.1) and thereby define the set of differential equations, there exists a \emph{unique} solution 
\begin{equation}
z_a(t) = \gamma_a(z^0; t^0, t), \ \ a = 1,m
\end{equation}
of (1.1) with the property
\begin{equation}
z_a(t^0) = \gamma_a(z^0; t^0, t^0) = z^0_a, \ \ a = 1,m.
\end{equation}
Let us also introduce a vector $\gamma\in {\mathbb{R}}^m$ with components $\gamma_1,\cdots,\gamma_m$.  With this notation (1.4) and (1.5) can be written more compactly in the form
\begin{equation}
z(t) = \gamma(z^0; t^0, t), 
\end{equation}
with 
\begin{equation}
z(t^0) = \gamma(z^0; t^0, t^0)= z^0.
\end{equation}

Now assume that $f$ is analytic (within some domain) in $z$ and $t$. By this we mean that the functions $f_a$ are analytic (within some domain) in the components of $z$ and the time $t$.  Then, according to a theorem of Poincar\'{e},  the solution quantities $z_a$ given by (1.4) (i.e., the components $\gamma_a$ of $\gamma$) will be analytic (again within some domain) in the components of the initial condition vector $z^0$ and the times $t^0$ and $t$ [1,2,3].  In vector notation, we say that $z$ is analytic in $z^0$, $t^0$, and $t$. 

Poincar\'{e} established this result on a case-by-case basis as needed using  {\em Cauchy's} method of {\em majorants}.  It is now more commonly established in general using {\em Picard iteration}, and appears as background material in many standard texts on ordinary differential equations [4,5].  It is also worth noting that Poincar\'{e}'s result can be generalized.  Suppose, for example, that the functions $f_a$ are only piece-wise continuous in $t$ but are analytic in $z$ for each value of $t$.  Then the solution will be piece-wise differentiable in $t$.  Remarkably, it will remain analytic in $z^0$ [6,7].  If, piece-wise, the functions $f_a$ have some order of differentiability in $t$ and analyticity in $z$, then the solution will have, piece-wise, one order more differentiability in $t$ and analyticity in $z^0$.  

Suppose that $z^d(t)$ is some given {\em design} solution to these equations, and we wish to study solutions in the vicinity of this solution.  That is, we wish to make expansions about this solution.  Introduce a vector of $m$ deviation variables $\zeta$ by writing
\begin{equation}
z = z^d + \zeta .
\end{equation}
Then the equations of motion (1.2) take the form
\begin{equation}
\dot{z}^d + \dot{\zeta} = f (z^d + \zeta ,t).
\end{equation}
We now assume that the right side of (1.2) is analytic about $z^d$.  Then we may write the relation
\begin{equation}
f (z^d + \zeta ,t) = f (z^d,t) + g(z^d,t,\zeta )
\end{equation}
where each component $g_a$ of $g$ has a Taylor expansion of the form
\begin{equation}
g_a(z^d,t,\zeta ) = \sum_r g_a^r(t) G_r(\zeta ).
\end{equation}
We also write (1.11) more compactly in the vector form
\begin{equation}
g(z^d,t,\zeta ) = \sum_r g^r(t) G_r(\zeta ).
\end{equation}
Here the $G_r(\zeta )$ are the various monomials in the $m$ components of $\zeta$ labeled by an {\em index} $r$ using some convenient labeling scheme, and the $g^r_a$ are (generally) time-dependent coefficients which we call {\em forcing terms}.  By construction, all the monomials $G_r(\zeta )$ occurring on the right sides of (1.11) and (1.12) have degree one or greater.  We note that the $g^r_a(t)$ are known once $z^d(t)$ is given.  

By assumption, $z^d$ is a solution of (1.2) and therefore satisfies the relations
\begin{equation}
\dot{z}^d = f(z^d,t)
\end{equation}
or, in terms of components,
\begin{equation}
\dot{z}^d_a = f_a(z^d,t).
\end{equation}
It follows that the deviation variables satisfy the equations of motion
\begin{equation}
{\dot{\zeta}}_a= g_a(z^d,t,\zeta ) = \sum_r g^r_a(t)G_r(\zeta )
\end{equation}
or, more compactly,
\begin{equation}
\dot{\zeta}= g(z^d,t,\zeta ) = \sum_r g^r(t)G_r(\zeta ).
\end{equation}
These equations are evidently generalizations of the usual first-degree (linear) variational equations, and will be called the {\em complete variational} equations.

Consider the solution to the complete variational equations with {\em initial} conditions $\zeta^i$ specified at some initial time $t^i$.  As described earlier, we expect that under suitable conditions this solution will be an analytic function of the initial conditions $\zeta^i$.  Also, since the right side of (1.16) vanishes when all the components of $\zeta$ vanish [all the monomials $G_r$ in (1.16) have degree one or greater], $\zeta (t) = 0$ is a solution to (1.16).  It follows that the solution to the complete variational equations has a Taylor expansion of the form
\begin{equation}
\zeta_a(t) = \sum_r h^r_a(t) G_r(\zeta^i)
\end{equation}
where the $h^r_a(t)$ are functions to be determined, and again all the monomials $G_r$ that occur have degree one or greater.  When the quantities $h^r_a(t)$ are evaluated at some {\em final} time $t^f$, (1.17) provides a representation of the transfer map ${\cal M}$ about the design orbit in the Taylor form
\begin{equation}
\zeta^f_a = \zeta_a(t^f) = \sum_r h^r_a(t^f) G_r(\zeta^i),
\end{equation}
or, more compactly,
\begin{equation}
\zeta^f = \sum_r h^r(t^f) G_r(\zeta^i)
\end{equation}
where each $h^r(t^f)$ is an array of $m$ entries consisting of the functions $h^r_a(t^f)$ with $a=1,m$.

The organization of this paper is as follows:  Section 2 treats, as an example, the Duffing equation and describes the properties of an associated stroboscopic transfer map $\cal{M}$.  These properties are obtained by numerical integration of the equations of motion.  This work is necessary to get the lay of the land.  For comparison, Section 3  studies some of the properties of the truncated Taylor maps ${\cal{M}}_n$ obtained by solving the variational equations numerically. There we will witness the remarkable fact that a truncated Taylor map approximation to $\cal M$ can reproduce the infinite period-doubling Feigenbaum cascade and associated strange attractor exhibited by the exact $\cal M$.  Truncation at order 3 already produces a map ${\cal{M}}_3$ with the right qualitative behavior, and agreement with exact numerical results improves as $n$ is increased beyond $n=3$.  Thus, in effect, we make a preliminary study of some convergence properties of truncated Taylor approximations to the Duffing equation stroboscopic map.  A final section provides a concluding summary and discussion.

\section{Duffing Equation Example}
\setcounter{equation}{0}
\subsection{Introduction}
As an example application, this section studies some aspects of the {\em Duffing} equation [8]. The behavior of the driven Duffing oscillator, like that of generic nonlinear systems, is enormously complicated.  Consequently, we will be able to touch only on some of the highlights of this fascinating problem.

Duffing's equation describes the behavior of a periodically driven damped {\em nonlinear} oscillator governed by the equation of motion
\begin{equation}
  \ddot{x} + a\dot{x} + bx + cx^3 = d \cos (\Omega t + \psi).
\end{equation} 
Here $\psi$ is an arbitrary phase factor that is often set to zero.  For our purposes it is more convenient to set 
\begin{equation}
\psi = \pi/2.
\end{equation}
Evidently any particular choice of $\psi$ simply results in a shift of the origin in time, and this shift has no physical consequence since the left side of (2.1) is independent of time.

 We assume $b,c>0$, which is the case of a positive hard spring restoring force.\footnote{Other authors consider other cases, particularly the `double well' case $b<0$ and $c>0$.}  We make these assumptions because we want the Duffing oscillator to behave like an ordinary harmonic oscillator when the amplitude is small, and we want the motion to be bounded away from infinity when the amplitude is large.  Then, by a suitable choice of time and length scales that introduces new variables $q$ and ${\tau}$, the equation of motion can be brought to the form
\begin{equation}
  \ddot{q} + 2\beta \dot{q} + q + q^3 = -\epsilon \sin \omega \tau ,
\end{equation}
where now a dot denotes $d/d\tau$ and we have made use of (2.2). In this form it is evident that there are 3 free parameters: $\beta$, $\epsilon$, and $\omega$.

\subsection{Stroboscopic Map}
While the Duffing equation is nonlinear, it does have the simplifying feature that the driving force is periodic with period
\begin{equation}
T = 2\pi /\omega .
\end{equation}
Let us convert (2.3) into a pair of first-order equations by making the definition
\begin{equation}
p = \dot{q},
\end{equation}
with the result
\[
\dot{q} = p,
\]
\begin{equation}
\dot{p} = -2\beta p - q - q^3 - \epsilon \sin \omega \tau .
\end{equation}
Let $q^0,p^0$ denote initial conditions at $\tau = 0$, and let $q^1,p^1$ be the final conditions resulting from integrating the pair (2.6) one full period to the time $\tau = T$.  Let $\cal{M}$ denote the transfer map that relates $q^1,p^1$ to $q^0,p^0$.  Then, using the notation $z=(q,p)$, we may write 
\begin{equation}
z^1 = {\cal{M}} z^0.
\end{equation}

Suppose we now integrate for a second full period to find $q^2,p^2$.  Since the right side of (2.6) is periodic, the rules for integrating from $\tau = T$ to $\tau = 2T$ are the same as the rules for integrating from $\tau = 0$ to $\tau = T$.  Therefore we may write
\begin{equation}
  z^2 = {\cal{M}}z^1 = {\cal{M}}^2z^0,
\end{equation}
and in general 
\begin{equation}
  z^{n+1} = {\cal{M}}z^n = {{\cal{M}}}^{n+1}z^0.
\end{equation}
We may regard the quantities $z^n$ as the result of viewing the motion in the light provided by a stroboscope that flashes at the times\footnote{Note that, with the choice (2.2) for $\psi$, the driving term described by the right side of (2.3) vanishes at the stroboscopic times $\tau^n$.}
\begin{equation}
  \tau^n = nT.
\end{equation}
Because of the periodicity of the right side of the equations of motion, the rule for sending $z^n$ to $z^{n+1}$ over the intervals between successive flashes is always the same, namely $\cal{M}$.  For these reasons $\cal{M}$ is called a \emph{stroboscopic map}.  Despite the explicit time dependence in the equations of motion, because of periodicity we have been able to describe the long-term motion by the repeated application of a single fixed map.

\subsection{Feigenbaum Diagram Overview}

One way to study a map and analyze its properties, in this case the Duffing stroboscopic map, is to find its fixed points.  When these fixed points are found, one can then display how they appear, move, and vanish as various parameters are varied.  Such a display is often called a {\em Feigenbaum} diagram.  This subsection will present selected Feigenbaum diagrams, including an infinite period doubling cascade and its associated strange attractor, for the stroboscopic map obtained by high-accuracy  numerical integration of the equations of motion (2.6).  They will be made by observing the behavior of fixed points as the driving frequency $\omega$ is varied.  For simplicity, the damping parameter will be held constant at the value $\beta=0.10$.  Various sample values will be used for the driving strength $\epsilon$.\footnote{Of course, one can also make Feigenbaum diagrams in which some other parameter, say $\epsilon$, is varied while the others, including $\omega$, are held fixed.}

\subsubsection{A Simple Feigenbaum Diagram}

Let us begin with the case of small driving strength.  When the driving strength is small, we know from energy considerations that the steady-state response will be small, and therefore the behavior of the steady-state solution will be much like that of the driven damped {\em linear} harmonic oscillator.  That is, for small amplitude motion, the $q^3$ term in (2.3) will be negligible compared to the other terms.  We also know that, because of damping, there will be only {\em one} steady-state solution, and therefore $\cal M$ has only {\em one} fixed point $z^f$ such that
\begin{equation}
{\cal M}z^f=z^f.
\end{equation}
(Here we use the super or subscript $f$ to denote {\em fixed} whereas in earlier sections it denoted {\em final}.)
Finally, again because of damping, we know that this fixed point is {\em stable}.  That is, if $\cal M$ is applied repeatedly to a point near $z^f$, the result is a sequence of points that approach ever closer to $z^f$.  For this reason a stable fixed point is also called an {\em attractor}.

Figure 1 shows the values of $q_f(\omega)$ for the case $\epsilon=0.150$, and Figure 2 shows $p_f(\omega)$.  In the figures the phase-space axes are labeled as $q_\infty$ and $p_\infty$ to indicate that what are being displayed are steady-state values reached after a large number of applications of $\cal M$.  As anticipated, we observe from Figures 1 and 2 that the response is much like the resonance response of the driven damped linear harmonic oscillator.\footnote{It was the desire for $q_\infty$ to exhibit a resonance-like peak as a function of $\omega$ that dictated the choice (2.2) for $\psi$.}  Note that the coefficient of $q$ in (2.3) is 1, and therefore at small amplitudes, where $q^3$ can be neglected, the Duffing oscillator described by (2.3) has a natural frequency near 1.  Correspondingly, Figure 1 displays a large response when the driving frequency has the value $\omega\simeq1$.  Observe, however, that the response, while similar, is not exactly like that of the driven damped linear harmonic oscillator.  For example, the resonance peak at $\omega\simeq1$ is slightly tipped to the right, and there is also a small peak for $\omega\simeq1/3$.

\begin{figure}[htp]
  \centering
  \includegraphics*[height=4.5in,angle=0]{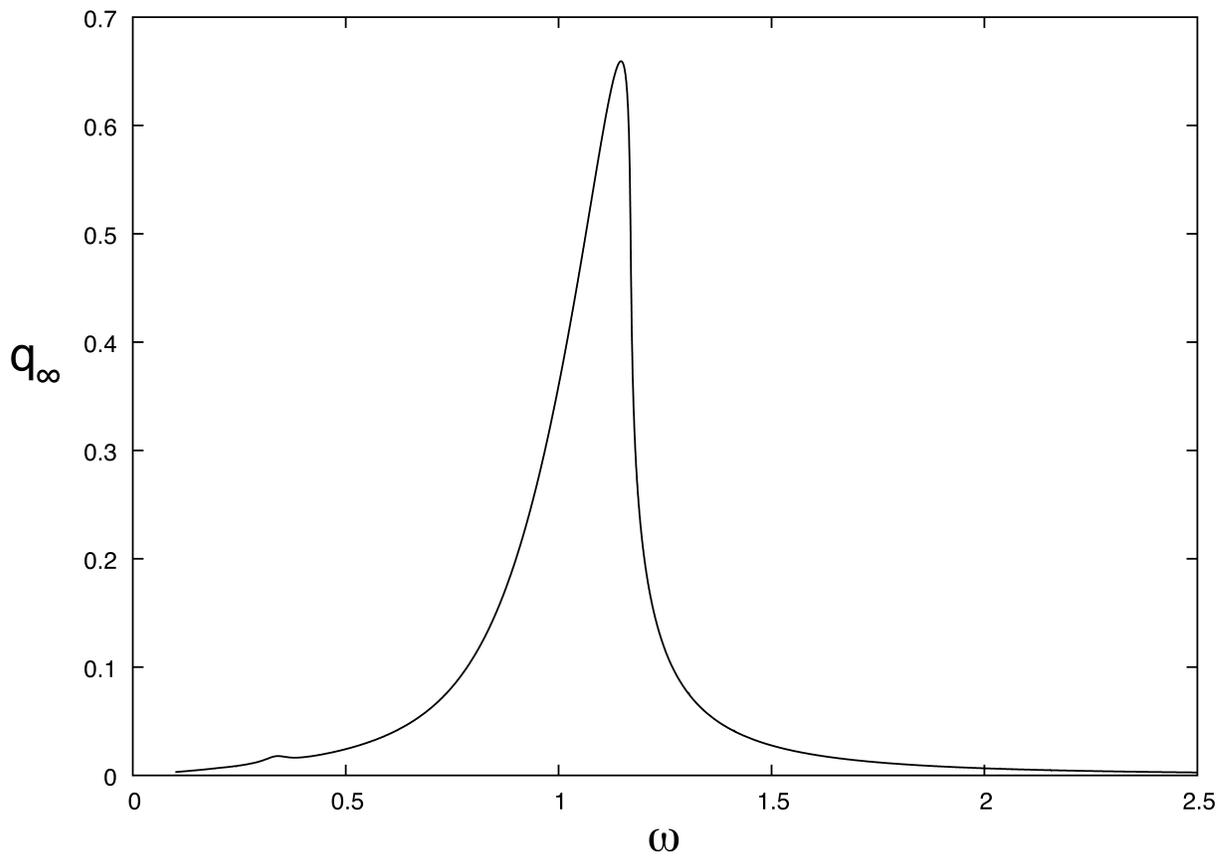}
  \caption{Feigenbaum diagram showing limiting values  $q_{\infty}$ as a function of $\omega$ (when $\beta = 0.1$ and $\epsilon = .15$) for the stroboscopic Duffing map.}
\end{figure}

\begin{figure}[htp]
  \centering
  \includegraphics*[width=0.7\textwidth]{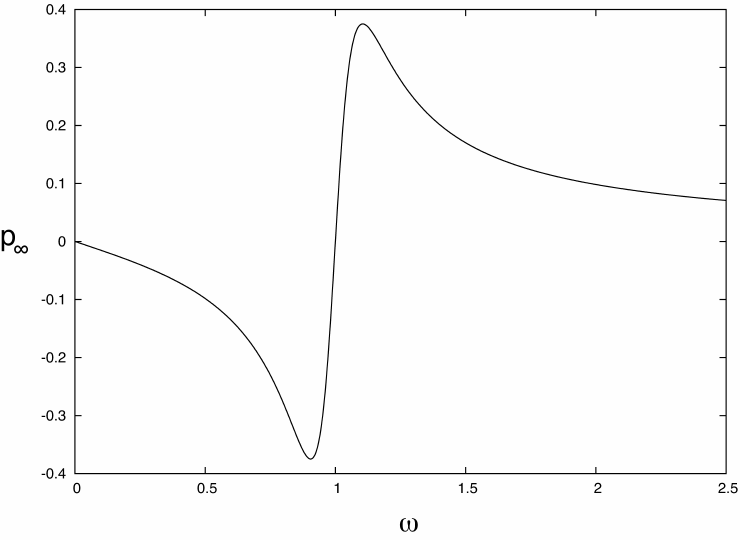}
  \caption{Feigenbaum diagram showing limiting values $p_{\infty}$ as a function
           of $\omega$ (when $\beta = 0.1$ and $\epsilon = .15$) for the
           stroboscopic Duffing map.}
\end{figure}

Our strategy for further exploration will be to increase the value of the driving strength $\epsilon$, all the while observing the stroboscopic Duffing map Feigenbaum diagram as a function of $\omega$.  We hasten to add the disclaimer that the driven Duffing oscillator displays an enormously rich behavior that varies widely with the parameter values $\beta$, $\epsilon$, $\omega$, and we shall be able to give a brief summary of only some of it.  Also, for brevity, we shall generally only display $q_f(\omega)$.

\subsubsection{Saddle-Node (Blue-Sky) Bifurcations}

Figure 3 shows the $q$ Feigenbaum diagram for the case of somewhat larger driving strength, $\epsilon=1.50$.  For this driving strength the resonance peak, which previously occurred at $\omega\simeq1$, has shifted to a higher frequency and taken on a more complicated structure.  There are now also noticeable resonances at lower frequencies, with the most prominent one being at $\omega\simeq1/2$. 

Examination of Figure 3 shows that for $\omega\le1.5$ there is a single stable fixed point whose trail is shown in black.  Then, as $\omega$ is increased, a pair of fixed points is born at $\omega\simeq1.8$.\footnote{Actually, in the analytic spirit of Poincar\'{e}, these fixed points also exist for smaller values of $\omega$, but are then complex.  They first become purely real, and therefore physically apparent, when $\omega\simeq1.8$.}  One of them is stable.  The other, whose trail as $\omega$ is varied is shown in red, is {\em unstable}.  That is, if ${\cal M}$ is applied repeatedly to a point near this fixed point, the result is a sequence of points that move ever farther away from the fixed point.  For this reason an unstable fixed point is also called a {\em repellor}.

This appearance of two fixed points out of nowhere is called a saddle-node {\em bifurcation} or a {\em blue-sky} bifurcation, and the associated Feigenbaum diagram is then sometimes called a bifurcation diagram.\footnote{Strictly speaking, a Feigenbaum diagram displays only the trails of stable fixed points while a bifurcation diagram displays the trails of all fixed points.}  The original stable fixed point persists as $\omega$ is further increased so that over some $\omega$ range there are 3 fixed points.  Then, as $\omega$ is further increased, the original fixed point and the unstable fixed point move until they meet and annihilate when  
$\omega\simeq2.6$.\footnote{Actually, they are not destroyed, but instead become complex and therefore disappear from view.}  This disappearance is called an inverse saddle-node or inverse blue-sky bifurcation.  Finally, for still larger $\omega$ values there is again only one fixed point, and it is stable. 

\begin{figure}[htp]
  \centering
  \includegraphics*[width=5in,angle=0]{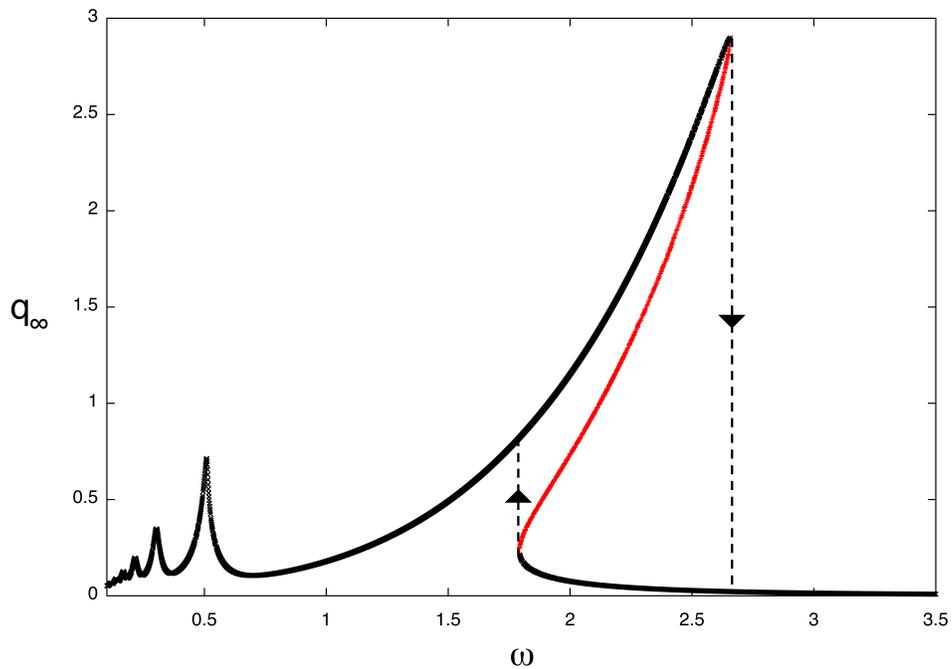}
  \caption{Feigenbaum/bifurcation diagram showing limiting values  $q_{\infty}$ as a function of $\omega$ (when $\beta = 0.1$ and $\epsilon = 1.5$) for the stroboscopic Duffing map.  Trails of the stable fixed points are shown in black.  Also shown, in red, is the trail of the unstable fixed point. Finally, jumps in the steady-state amplitude are illustrated by vertical dashed lines at $\omega \simeq 1.8$ and $\omega \simeq 2.6$.}
\end{figure}

We remark that the appearance and disappearance of stable-unstable fixed-point pairs, as $\omega$ is varied, has a striking dynamical consequence.  Suppose, for example in the case of Figure 3, that the driving frequency $\omega$ is below the value $\omega \simeq 1.8$ where the saddle-node bifurcation occurs.  Then there is only one fixed point, and it is attracting.  Now suppose $\omega$ is {\em slowly} increased.  Then, since the fixed-point solution is attracting, the solution for the slowly increasing $\omega$ case will remain near this solution.  See the upper black trail in Figure 3.  This ``tracking" will continue until $\omega$ reaches the value  $\omega \simeq 2.6$ where the inverse saddle-node bifurcation occurs.  At this value the fixed point being followed disappears.  Consequently, since the one remaining fixed point is also an attractor, the solution evolves very quickly to that of the remaining fixed point.  It happens that the oscillation amplitude associated with this fixed point is much smaller, and therefore there appears to be a sudden jump in oscillation amplitude to a smaller value.  Now suppose $\omega$ is slowly decreased from a value above the value $\omega \simeq 2.6$ where the inverse saddle-node bifurcation occurs.  Then the solution will remain near that of the fixed point lying on the bottom black trail in Figure 3.  This tracking will continue until $\omega$ reaches the value $\omega \simeq 1.8$ where the fixed point being followed disappears.  Again, since the remaining fixed point is attracting, the solution will now evolve to that of the remaining fixed point.  The result is a jump to a larger oscillation amplitude.  Evidently the steady-state oscillation amplitude exhibits {\em hysteresis} as $\omega$ is slowly varied back and forth over an interval that begins below the value where the first saddle-node bifurcation occurs and ends at a value above that where the inverse saddle-node bifurcation occurs.

\subsubsection{Pitchfork Bifurcations}

Let us continue to increase $\epsilon$. Figure 4 shows that a qualitatively new feature appears when $\epsilon$ is near 2.2: a {\em bubble} is formed between the major resonant peak (the one that has saddle-node bifurcated) and the subresonant peak immediately to its left.  To explore the nature of this bubble, let us make $\epsilon$ still larger, which, we anticipate, will result in the bubble becoming larger. Figure 5 shows the Feigenbaum diagram in the case $\epsilon = 5.5$.  Now the major resonant peak and the subresonant peak have moved to larger $\omega$ values. Correspondingly, the bubble between them has also moved to larger $\omega$ values.  Moreover, it is larger, yet another smaller bubble has formed, and the subresonant peak between them has also undergone a saddle-node bifurcation.  For future use, we will call the major resonant peak the {\em first} or {\em leading} saddle-node bifurcation, and we will call the subresonant peak between the two bubbles the {\em second} saddle-node bifurcation, etc.  Also, we will call the bubble just to the left of the first saddle-node bifurcation the {\em first} or {\em leading} bubble, and the next bubble will be called the {\em second} bubble, etc.

We also note that three short trails have appeared in Figure 5 just to the right of $\omega=4$.  They correspond to a period-three bifurcation followed shortly thereafter by an inverse bifurcation.   Actually, much closer examination shows that there are six trails consisting of three closely-spaced pairs.  Each pair comprises a stable and an unstable fixed point of the map ${\cal {M}}^3$.  They are not fixed points of $\cal M$ itself, but rather are sent into each other in cyclic fashion under the action of $\cal M$.

\begin{figure}[htp]
  \centering
  \includegraphics*[width=0.75\textwidth]{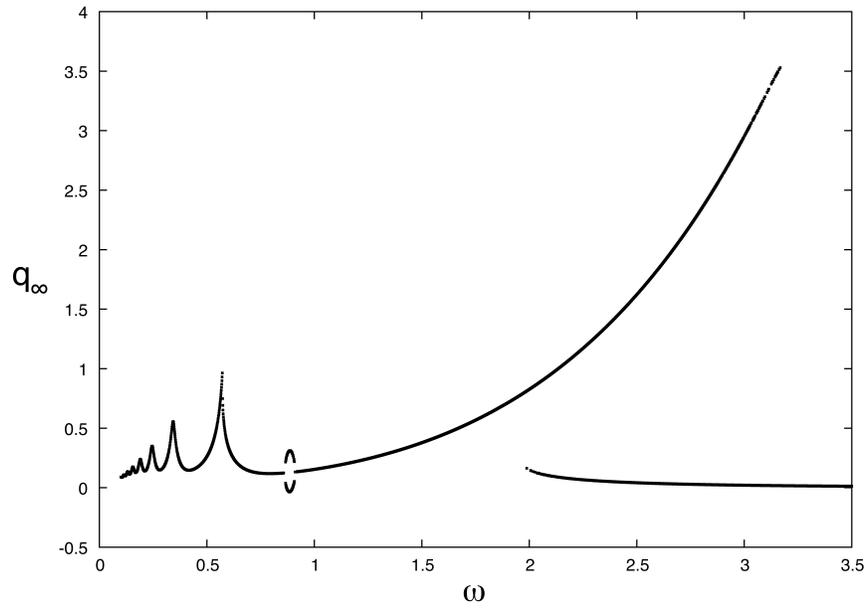}
  \caption{Feigenbaum diagram showing limiting values  $q_{\infty}$ as a function of $\omega$ (when $\beta = 0.1$ and $\epsilon = 2.2$) for the stroboscopic Duffing map. It displays that a bubble has now formed at $\omega\approx.8$.}
\end{figure}

\begin{figure}[htp]
  \centering
  \includegraphics*[width=0.75\textwidth]{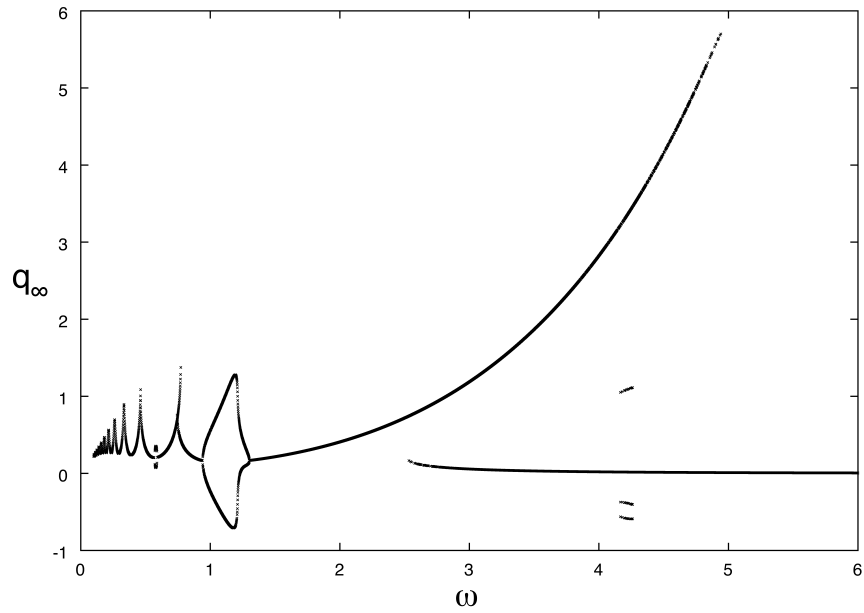}
  \caption{Feigenbaum diagram showing limiting values  $q_{\infty}$ as a function of $\omega$ (when $\beta = 0.1$ and $\epsilon = 5.5$) for the stroboscopic Duffing map. The first bubble has grown, a second smaller bubble has formed to its left, and the subresonant peak between them has saddle-node bifurcated to become the second saddle-node bifurcation.}
\end{figure}

Figure 6 shows the larger (leading) bubble in Figure 5 in more detail and with the addition of red lines indicating the trails of unstable fixed points. It reveals that the bubble describes the {\em simultaneous} bifurcation of a single fixed point into three fixed points.  Two of these fixed points are stable and the third, whose $q$ coordinate as a function of $\omega$ are shown as a red line, is unstable.  What happens is that, as $\omega$ is increased, a {\em single} stable fixed point becomes a {\em triplet} of fixed points, two of which are stable and one of which is unstable. This is called a {\em pitchfork} bifurcation. Then, as $\omega$  is further increased, these three fixed points again merge, in an inverse pitchfork bifurcation, to form what is again a single stable fixed point. 

\begin{figure}[htp]
  \centering
   \includegraphics*[width=5in,angle=0]{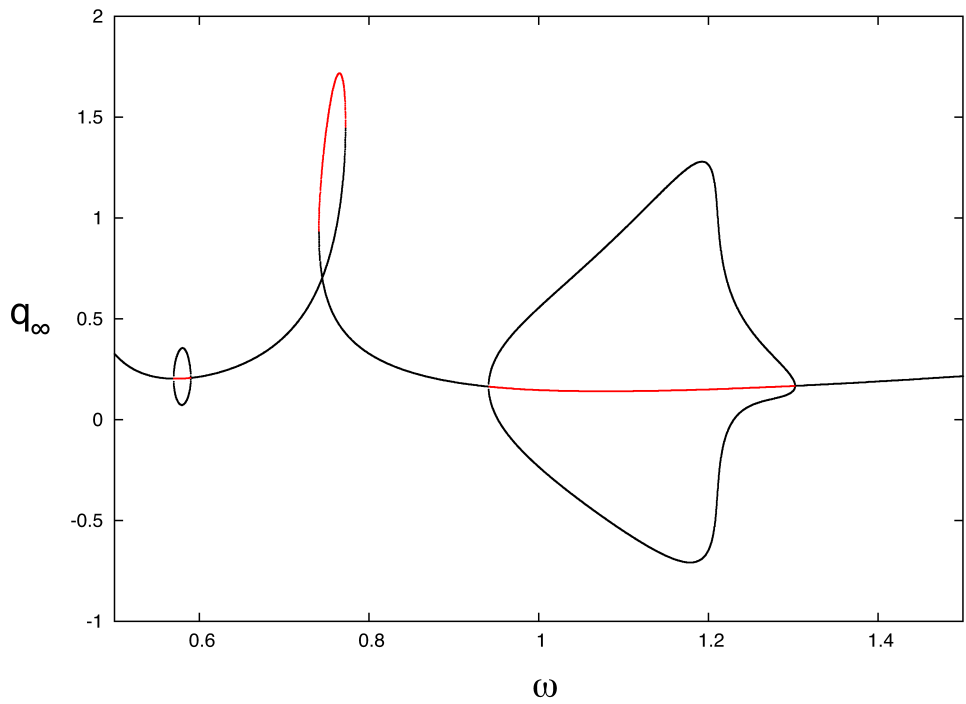}
  \caption{An enlargement of Figure 5 with the addition of red lines indicating the trails of unstable fixed points.}
\end{figure}

\newpage
\subsubsection{A Plethora of Bifurcations and Period Doubling Cascades}

We end our numerical study of the Duffing equation by increasing $\epsilon$ from its earlier value $\epsilon=5.5$ to much larger values.  First we will set $\epsilon=22.125$.  Based on our experience so far, we might anticipate that the Feigenbaum diagram would become much more complicated.  That is indeed the case.  Figure 7 displays  $q_\infty$  when $\beta=0.1$ and $\epsilon=22.125$, as a function of $\omega$,  for the range $\omega \in (0,12)$. Evidently the behavior of the attractors for the stroboscopic Duffing map, which is what is shown in Figure 7, is extremely complicated.  There are now a great many fixed points both of $\cal M$ itself and various powers of $\cal M$.  For small values of $\omega$, and as in Figures 3 through 5, there  are many resonant peaks and numerous saddle-node and pitchfork bifurcations.  For larger values of $\omega$ there are more complicated bifurcations.   In this figure, and some subsequent figures, the coloring scheme is chosen to guide the eye in following bifurcation trees with colors changing when the period changes.  Points with period one are colored red and points of very high or no discernible period are colored black.

Of particular interest to us are the two areas around $\omega=.8$ and $\omega=1.25$.  They contain what has become of the first two bubbles in Figure 6, and are shown in greater magnification in Figure 8.  What has happened is that bubbles have formed within bubbles, and bubbles have formed within these bubbles, etc. to produce a cascade.  However, these interior bubbles are not the result of pitchfork bifurcations, but rather the result of {\em period-doubling} bifurcations.  For example, the bifurcation that creates the first bubble at $\omega\simeq1.2$ is a pitchfork bifurcation.  But the successive bifurcations within the bubble are period-doubling bifurcations.  In a period-doubling bifurcation a fixed point that is initially stable becomes unstable as $\omega$ is increased.  When this happens, simultaneously two stable fixed points of ${\cal {M}}^2$ are born.  They are not fixed points of $\cal M$ itself, but rather are sent into each other under the action of $\cal M$.  Hence the name ``period doubling".  The map $\cal M$ must be applied twice to send such a fixed point back into itself.  In the next period doubling, fixed points of ${\cal {M}}^4$ are born, etc.  However we note that, as $\omega$ increases,  the sequence of period-doubling bifurcations only occurs a {\em finite} number of times and then undoes itself.

Remarkably, when $\epsilon$ is just somewhat larger, {\em infinite} sequences of period doubling cascades can occur.  Figure 9 shows what happens when $\epsilon=25$.

\begin{figure}[htp]
  \centering
    \includegraphics*[height=5in,angle=90]{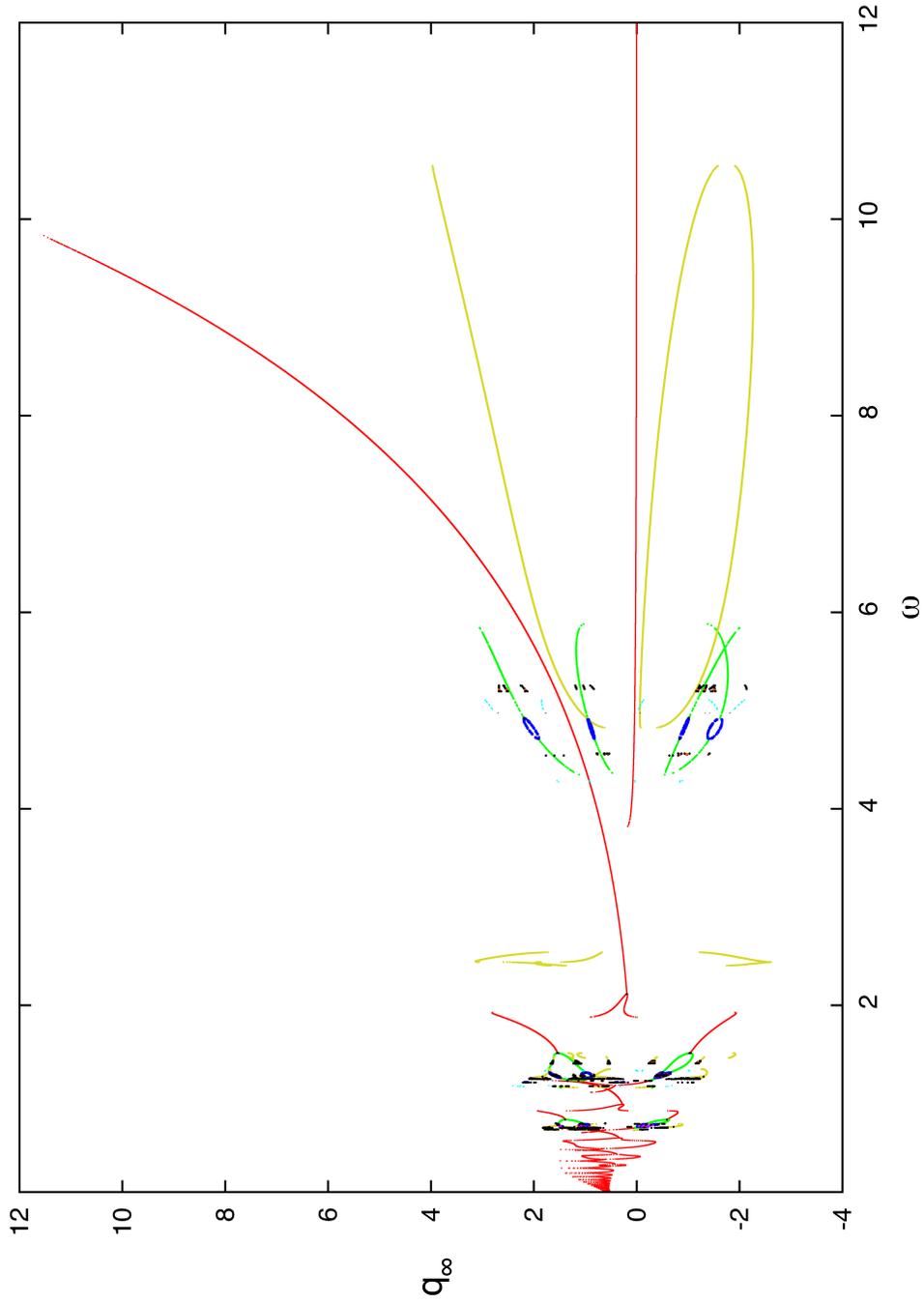}
  \caption{Feigenbaum diagram showing limiting values $q_{\infty}$ as a function of $\omega$ (when $\beta = 0.1$ and $\epsilon = 22.125$) for the stroboscopic Duffing map.}
\end{figure}

\begin{figure}[htp]
  \centering
  \includegraphics*[height=4in,angle=0]{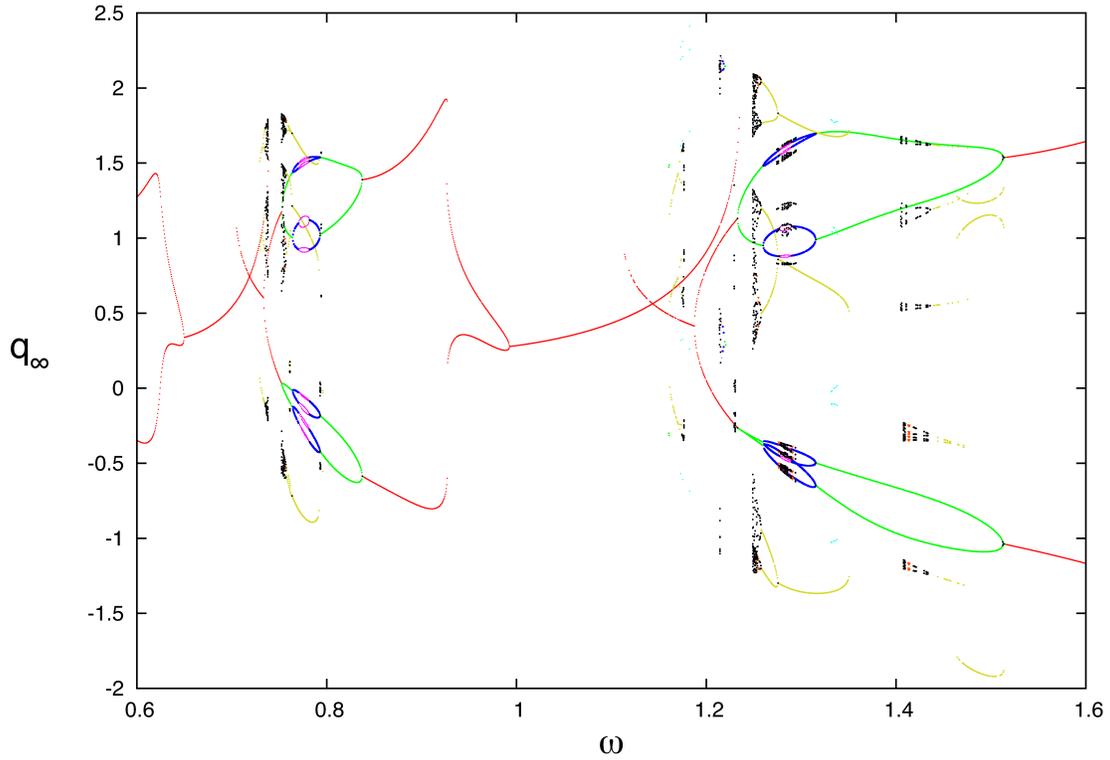}

  \caption{Enlarged portion of the Feigenbaum diagram of Figure 7 displaying  limiting values $q_{\infty}$ as a function of $\omega$ (when $\beta = 0.1$ and $\epsilon = 22.125$) for the stroboscopic Duffing map.  It shows part of the  first bubble at the far right, the second bubble, and part of a third bubble at the far left. Examine the first and second bubbles.  Each initially consists of two stable period-one fixed points.  Each also contains, as $\omega$ is increased, the beginnings of  period-doubling cascades. These cascades do not complete, but rather cease and then undo themselves by successive mergings as $\omega$ is further increased.  The final result is again a pair of stable period-one fixed points.  There are also many higher-period fixed points and their associated cascades.}
\end{figure}

\begin{figure}[htp]
  \centering
  \includegraphics*[height=5in,angle=90]{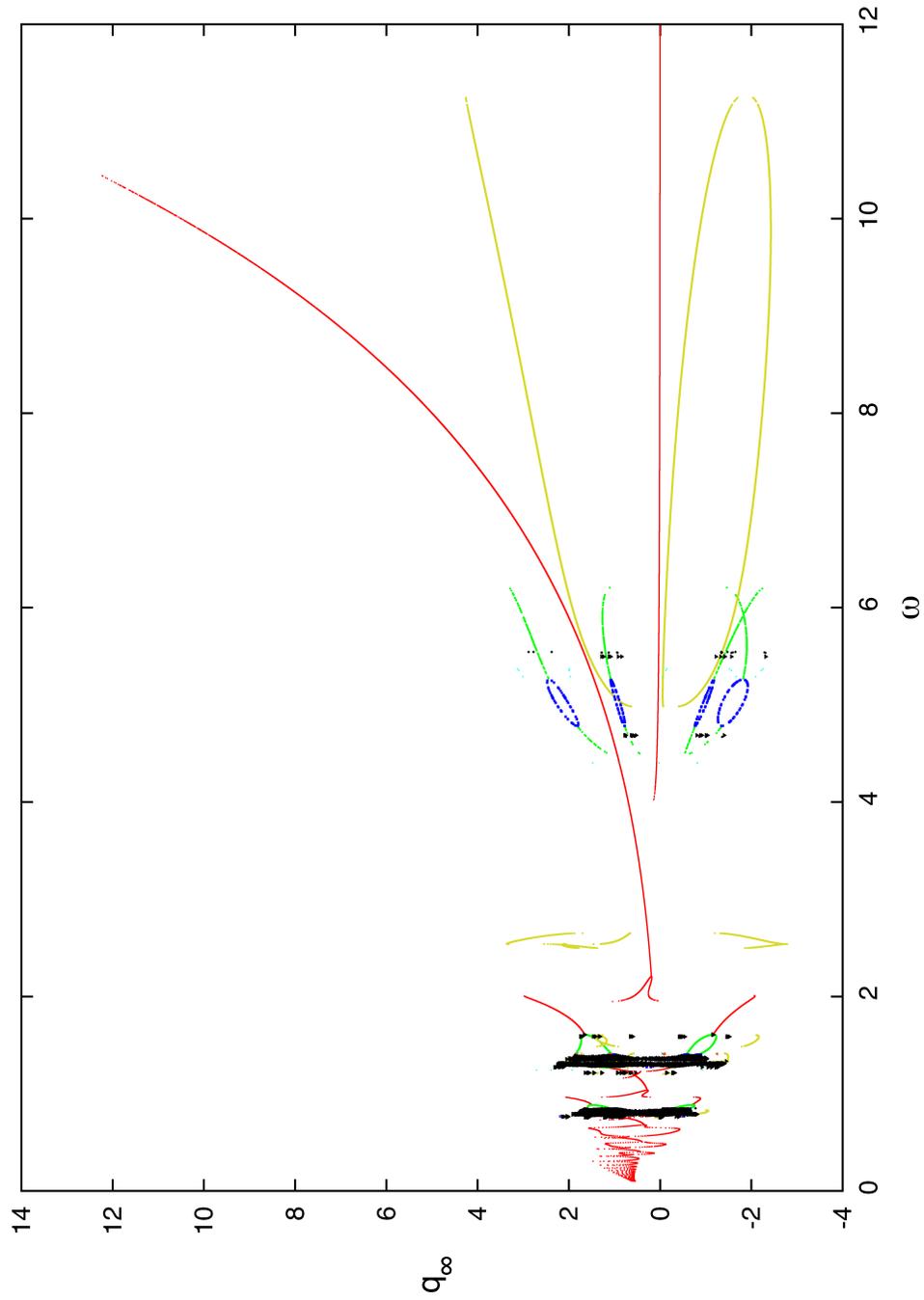}
  \caption{Feigenbaum diagram showing limiting values $q_{\infty}$ as a function of $\omega$ (when $\beta = 0.1$ and $\epsilon = 25$) for the stroboscopic Duffing map.}
\end{figure}

\newpage
\subsubsection{More Detailed View of Infinite Period Doubling Cascades}

To display the infinite period doubling cascades in more detail, Figure 10 shows an enlargement of part of Figure 9.  And Figures 11 and 12 show successive enlargements of parts of the first bubble in Figure 9.  From Figure 10 we see that the first bubble forms as a result of a pitchfork bifurcation just to the right of $\omega=1.2$, and from Figures 11 and 12 we see that the first period doubling bifurcation occurs in the vicinity of $\omega=1.268$.  From Figure 12 it is evident that successive period doublings occur an infinite number of times to ultimately produce a chaotic region when $\omega$ exceeds $\omega\simeq1.29$.

\begin{figure}[htp]
  \centering
  \includegraphics*[height=4in,angle=0]{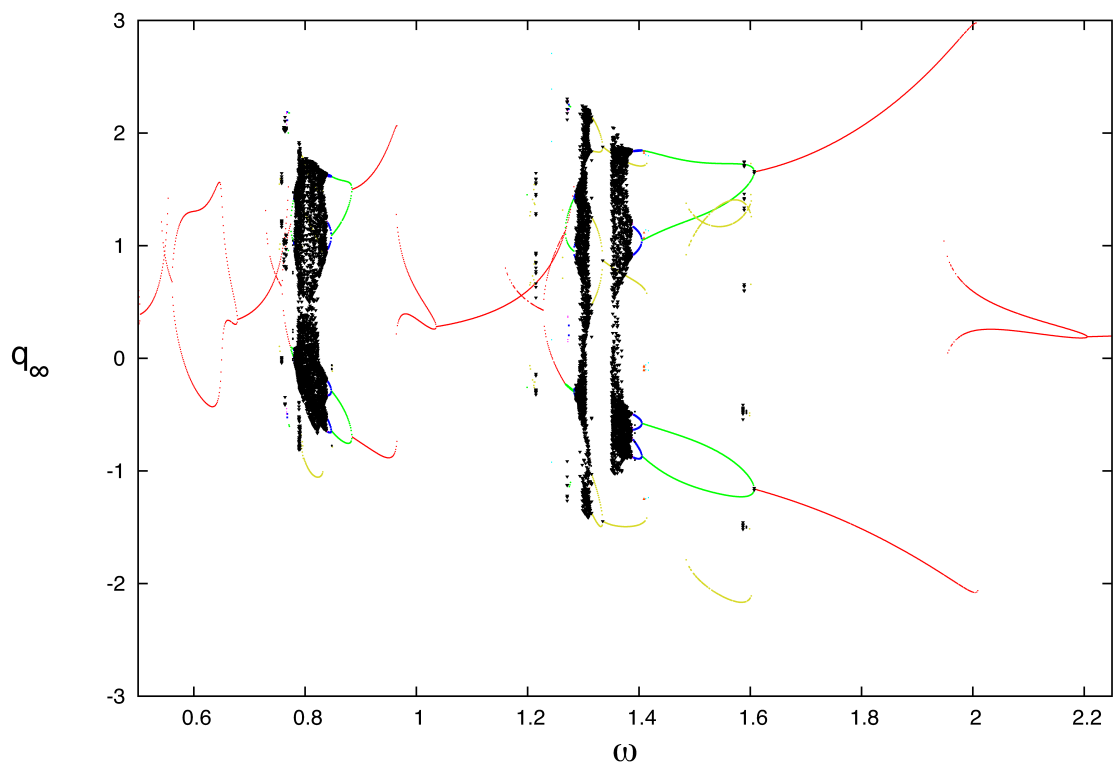}
  \caption{Enlargement of a portion Figure 9 showing the first, second, and third bubbles. The period-doubling cascades in each of the first and second bubbles complete. Then they undo themselves as $\omega$ is further increased. There is no period doubling in the third bubble when $\epsilon=25$.}
\end{figure}

\begin{figure}[htp]
  \centering
  \includegraphics*[height=3.5in]{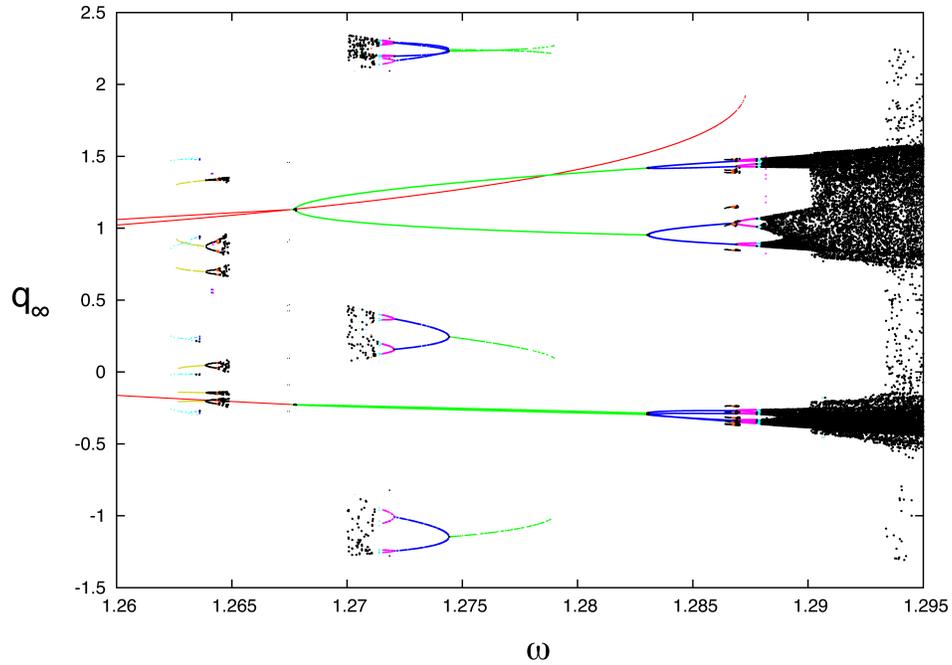}
  \caption{Detail of part of the first bubble in Figure 10  showing upper and lower infinite period-doubling cascades. Part of the trail of the stable fixed point associated with the second saddle-node bifurcation accidentally appears to overlay the upper period doubling bifurcation. Finally, there are numerous cascades followed by successive mergings  associated with higher-period fixed points.}
\end{figure}

\begin{figure}[htp]
  \centering
  \includegraphics*[height=3.5in]{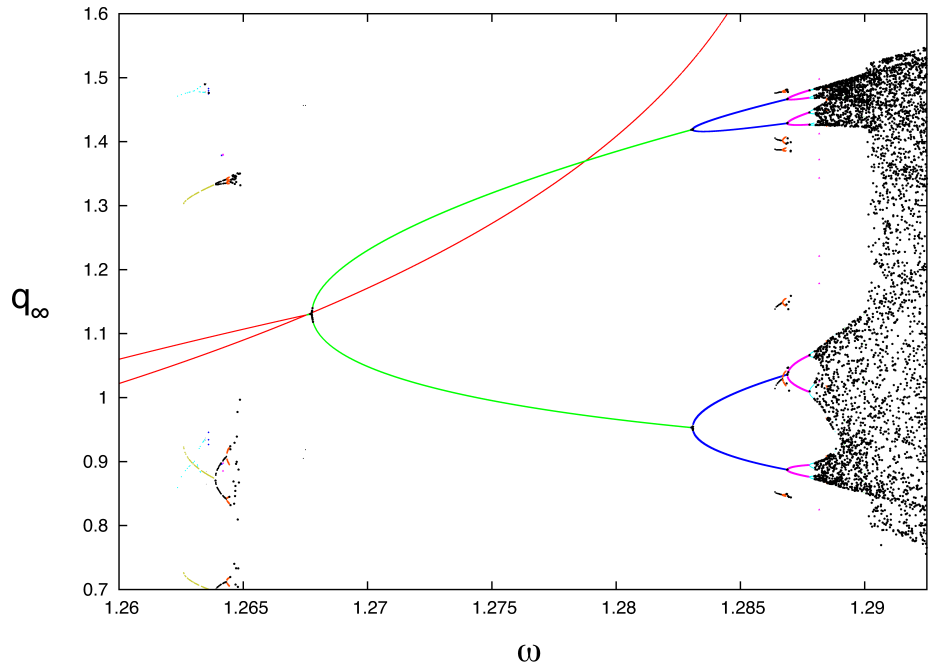}
  \caption{Detail of part of the upper cascade in Figure 11 showing an infinite period-doubling cascade, followed by chaos, for what was initially a stable period-one fixed point.}
\end{figure}

\subsubsection{Strange Attractor}

As evidence that the behavior in this region is chaotic, Figures 13 and 14 show portions of the {\em full} phase space, the $q,p$ plane, when $\omega=1.2902$.  Note the evidence for fractal structure.  The points appear to lie on a {\em strange attractor}.

\begin{figure}[htp]
  \centering
   \includegraphics*[height=3.5in]{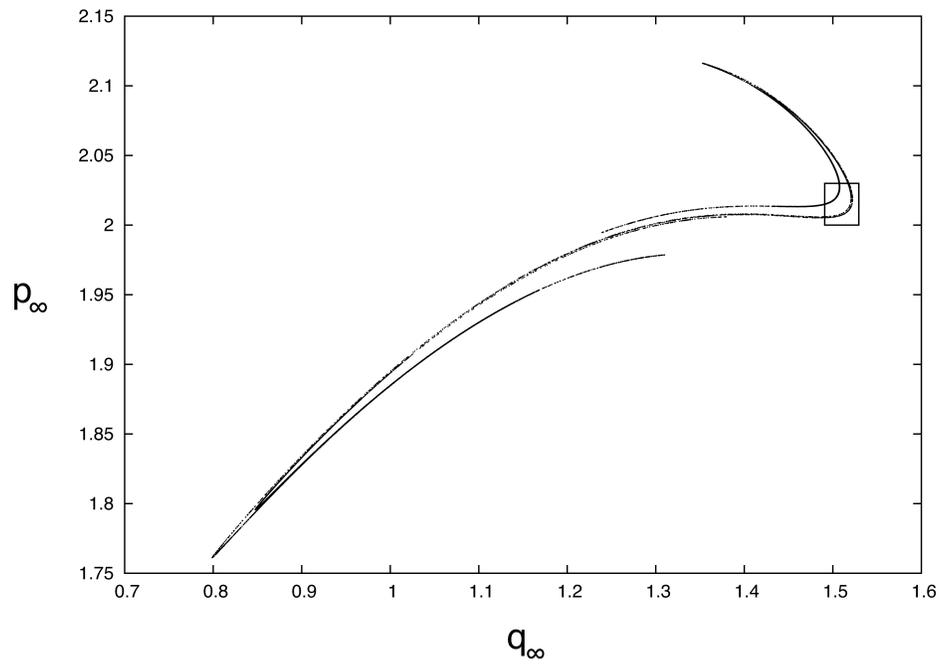}
  \caption{Limiting values of $q_{\infty},p_{\infty}$ for the stroboscopic Duffing map when $\omega=1.2902$ (and $\beta=.1$ and $\epsilon=25$). They appear to lie on a   strange attractor.}
\end{figure}

\begin{figure}[htp]
  \centering
  \includegraphics*[height=3.5in]{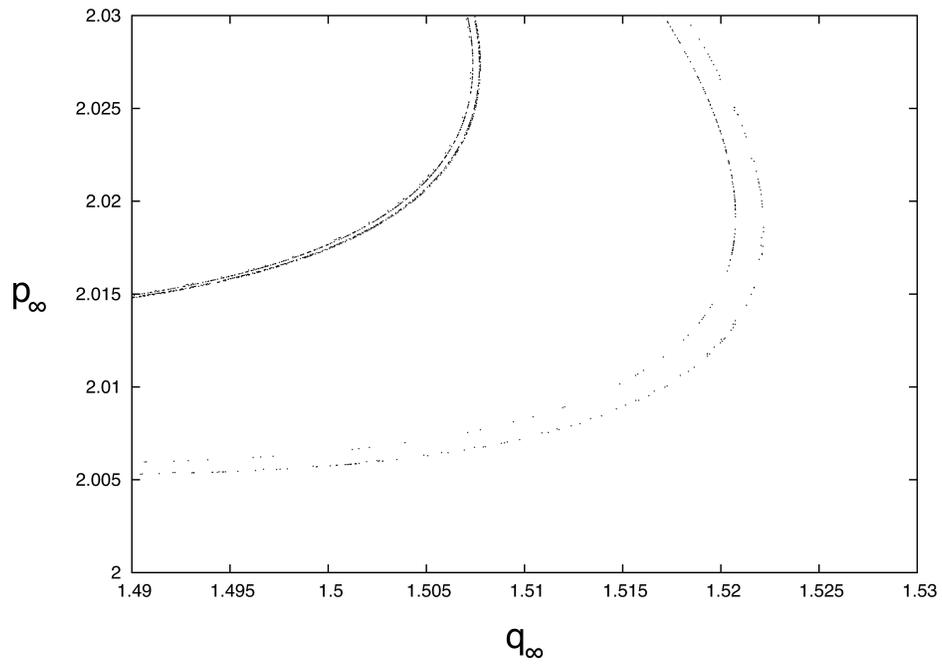}
  \caption{Enlargement of boxed portion of Figure 13 illustrating the
           beginning of self-similar fractal structure.}
\end{figure}

\newpage
\section{Polynomial Approximations to Duffing Stroboscopic Map}
\setcounter{equation}{0}

In this section we will compare the behavior of the exact map $\cal{M}$ with polynomial (truncated Taylor series) maps ${\cal{M}}_n$ obtained by integrating the complete variational equations.  In particular, we will make this comparison in the vicinity of the infinite period doubling cascade displayed in Figures 12 through 14 since duplicating the intricate behavior in this vicinity would seem to be particularly challenging.  As an additional challenge, we will employ polynomial maps that involve expansions in the parameter $\omega$ as well.  This is done by introducing an additional deviation variable $\zeta_3$ by writing
\begin{equation}
\omega=\omega^d+\zeta_3
\end{equation}
and augmenting the complete variational equations by the equation
\begin{equation}
\dot{\zeta}_3=0.
\end{equation}
Here $\omega^d$ is the value of $\omega$ for the design solution.  By an application of Poincar\'{e}'s theorem in the case of parameter dependence, the solution to \begin{equation}
\dot{z}= f(z,t,\lambda),
\end{equation}
where $\lambda$ denotes a set of parameters, is analytic in any parameter providing $f$ is analytic in that parameter.  The dependence of the right sides of (2.6) is manifestly analytic in $\omega$ and we therefore know that the solution to (2.6) will be analytic in $\zeta_3$.  For further discussion of this point, and a description of how the complete variational equations can be integrated to obtain the desired truncated Taylor map, see the references [9,10].

\subsection{Performance of High-Order Polynomial Approximation}

Let ${\cal{M}}_8$ denote the $8^{\rm{th}}$-order polynomial map (with parameter dependence) approximation to the stroboscopic Duffing map $\cal M$.  Provided the relevant phase-space region is not too large, we have found that ${\cal{M}}_8$  reproduces all the features, described in Subsection 2.3, of the exact map [10].  (The phase-space region must lie within the convergence domain of the Taylor expansion.)  This reproduction might not be too surprising in the cases of elementary bifurcations such as saddle-node and pitchfork bifurcations.  What is more fascinating, as we will see, is that  ${\cal{M}}_8$ also reproduces the infinite period doubling cascade and associated strange attractor displayed in Figures 12 through 14.

\subsubsection{Infinite Period Doubling Cascade}

Figure 15 shows the partial Feigenbaum diagram for the map ${\cal{M}}_8$ in the case that $\beta=0.1$ and $\epsilon=25$.  The Taylor expansion is made about the point indicated by the {\em black dot}.  This point (the initial conditions for the design solution) has the coordinates
\begin{equation}
q_{\rm{bd}}=1.26082,{\;}p_{\rm{bd}}=2.05452,{\;}\omega_{\rm{bd}}=1.285.
\end{equation}
It was selected to be an unstable fixed point of $\cal M$, but that is not essential.  Any nearby expansion point would have served as well.
Note the remarkable resemblance of Figures 12 and 15.

We have referred to Figure 15 as a {\em partial} Feigenbaum diagram because it shows only $q_\infty$ and not $p_\infty$.  In order to give a complete picture, Figure 16 displays them both.

\begin{figure}[htp]
  \centering
  \includegraphics*[width=4.5in]{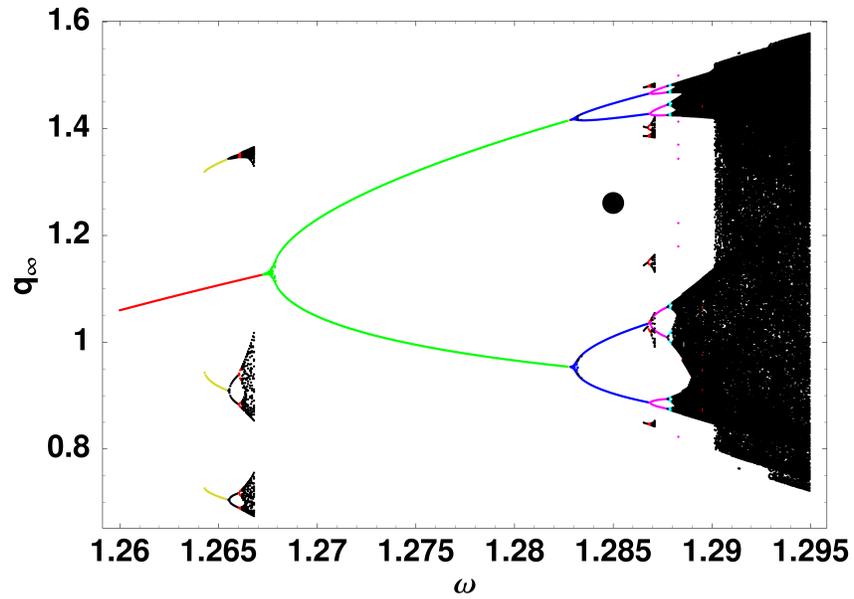}
  \caption{Partial Feigenbaum diagram for the map  ${\cal {M}}_8$. The black dot marks the point about which ${\cal M}$ is expanded to yield ${\cal {M}}_8$}.
\end{figure}

\begin{figure}[htp]
  \centering
  \includegraphics*[width=4.5in]{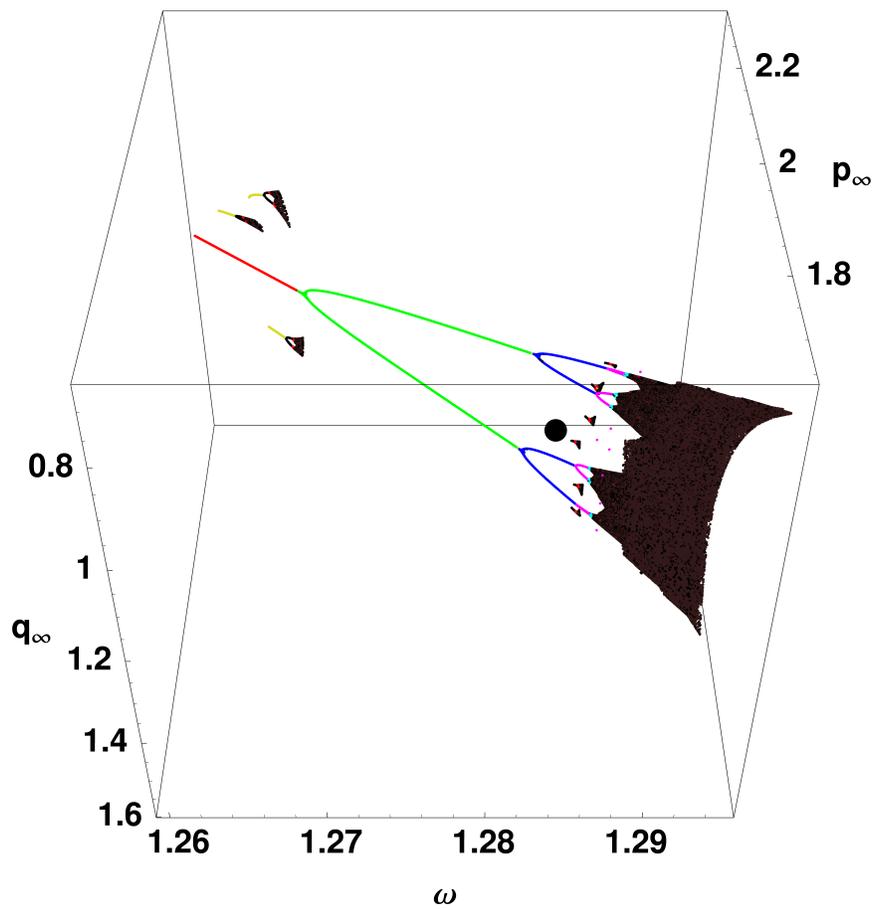}
  \caption{Full Feigenbaum diagram for the map  ${\cal {M}}_8$. The black dot again marks the expansion point.}
\end{figure}

\newpage
\subsubsection{Strange Attractor}

As displayed in Figures 17 through 20 ${\cal {M}}_8$, like $\cal M$,  appears to have a strange attractor.  Note the remarkable agreement between Figures 13 and 14 for $\cal M$ and their ${\cal {M}}_8$ counterparts, Figures 17 and 18.  In the case of ${\cal {M}}_8$ we have been able to obtain additional {\em enlargements}, Figures 19 and 20, further illustrating a self-similar fractal structure.  Analogous  figures are more difficult to obtain for the exact map $\cal M$ due to the excessive numerical integration time required.  By contrast the map ${\cal {M}}_8$, because it is a simple polynomial, is easy to evaluate repeatedly.
 
\begin{figure}[htp]
  \centering
  \includegraphics*[height=4.5in]{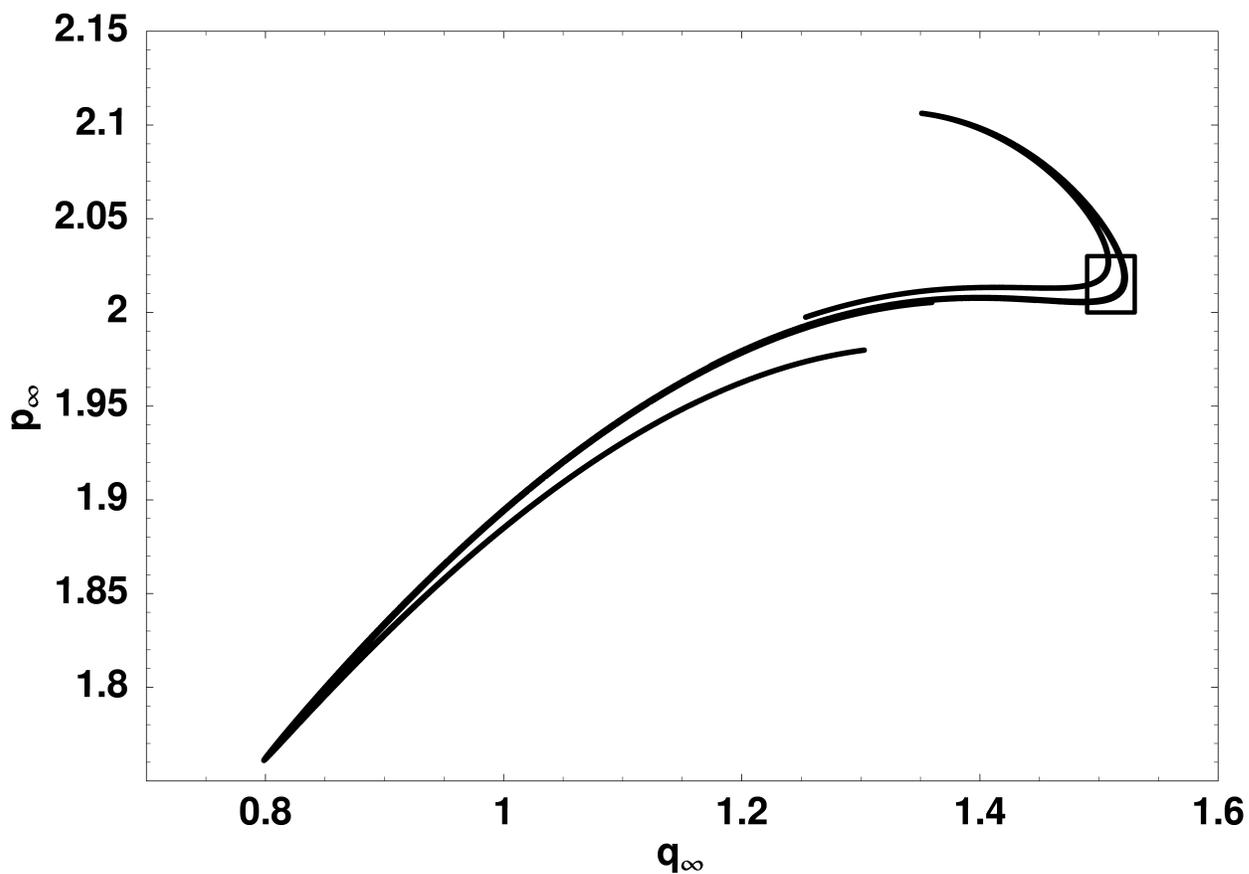}
  \caption{Limiting values of $q_{\infty},p_{\infty}$ for the map ${\cal {M}}_8$ when $\omega=1.2902$. They appear to lie on a   strange attractor.}
\end{figure}

\begin{figure}[htp]
  \centering
  \includegraphics*[width=5.5in]{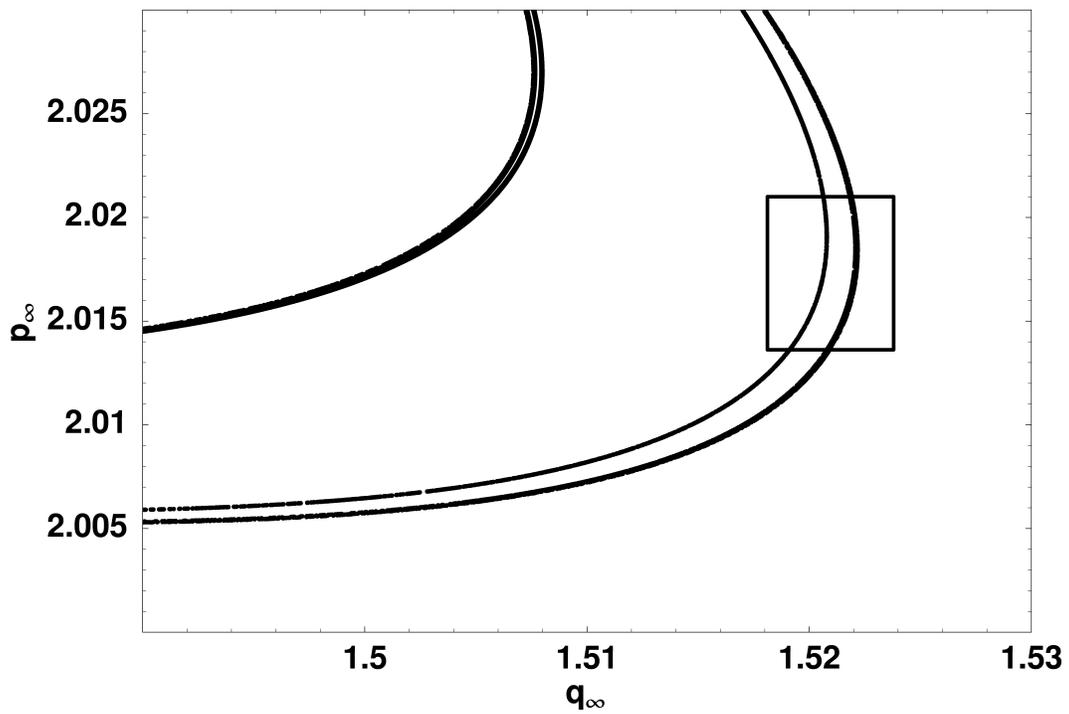}
  \caption{Enlargement of boxed portion of Figure 17 illustrating the
           beginning of self-similar fractal structure.}
\end{figure}

\begin{figure}[htp]
  \centering
  \includegraphics*[height=5.5in]{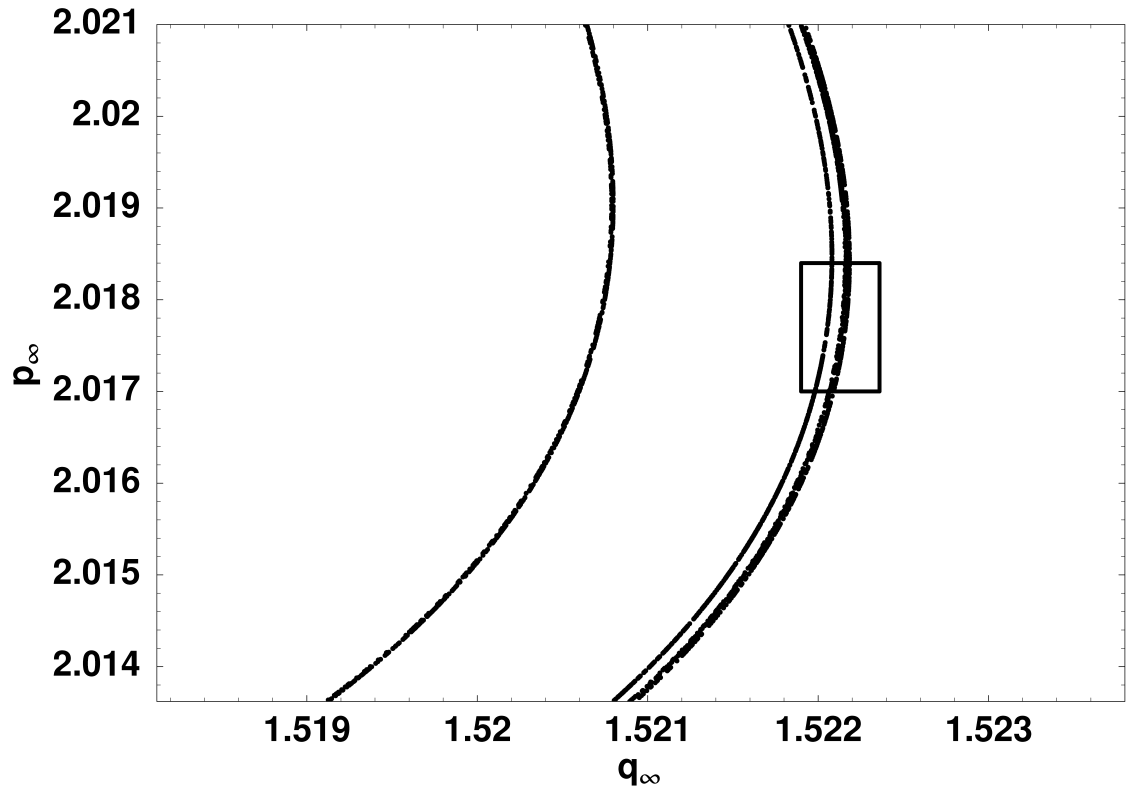}
  \caption{Enlargement of boxed portion of Figure 18 illustrating the
           continuation of self-similar fractal structure.}
\end{figure}

\begin{figure}[htp]
  \centering
  \includegraphics*[height=5.5in]{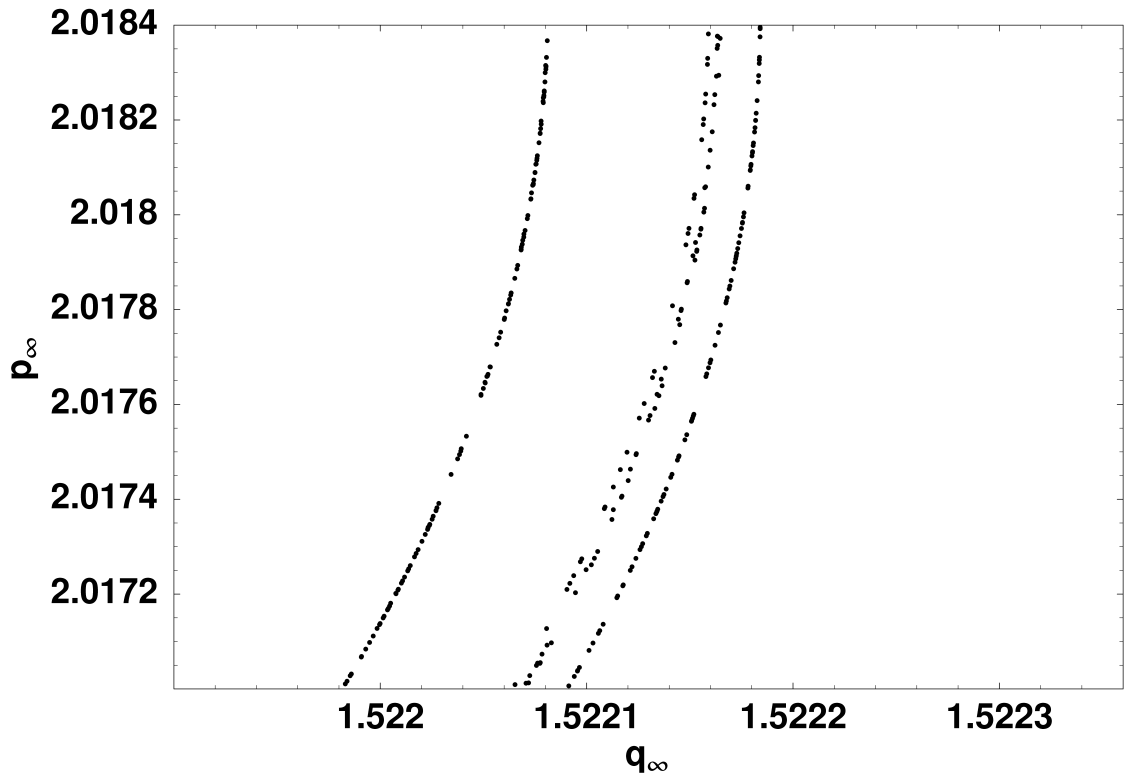}
  \caption{Enlargement of boxed portion of Figure 19 illustrating the
           further continuation of self-similar fractal structure.}
\end{figure}

\subsection{Performance of Lower-Order Polynomial Approximations}
We close this section with illustrations of the performances of ${\cal{M}}_3$ and ${\cal{M}}_5$, third and fifth order polynomial approximations (including parameter dependence) to the exact map $\cal{M}$.  All expansions are made about the point (3.4).  Comparison of these performances gives some feeling for the convergence properties of the Taylor approximation to $\cal{M}$.

\subsubsection{Performance of ${\cal{M}}_3$}

Figure 21 shows the ${\cal{M}}_3$ counterpart to  
Figure 15 produced using ${\cal{M}}_8$.  Evidently the qualitative features of the period doubling cascade are the same.  Also, we have found that there is not qualitative agreement if ${\cal{M}}_2$ is used.  We conjecture that generically third-order information is necessary and sufficient to obtain qualitative agreement for a period doubling cascade arising from what once was a period-one fixed point.  

Note also that  ${\cal{M}}_3$ does not reproduce the three features near $\omega=1.265$ seen in Figure 12 for the exact $\cal{M}$ and in Figure 15 for ${\cal{M}}_8$.  We have found that these features first appear at $n=5$.  They belong to what was initially a period-three fixed point for $\cal{M}$.

Figures 22 and 23 show the ${\cal {M}}_3$ counterparts to Figures 17
and 18 produced using ${\cal {M}}_8$. Evidently there is qualitative
agreement. The attractors in Figures 22 and 17 look similar. And, when
enlarged, both show evidence of fractal structure. Compare Figures 23
and 18.

\begin{figure}[htp]
  \centering
  \includegraphics*[width=4.5in]{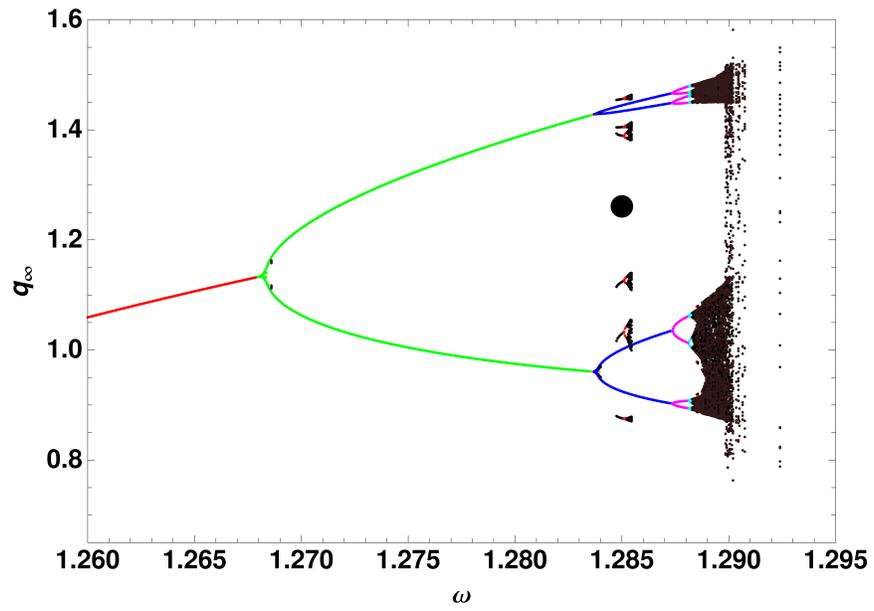}
  \caption{
Partial Feigenbaum diagram for the map  ${\cal {M}}_3$. The black dot marks the point about which ${\cal M}$ is expanded to yield ${\cal {M}}_3$}.
\end{figure}

\begin{figure}[htp]
  \centering
  \includegraphics*[height=4.5in]{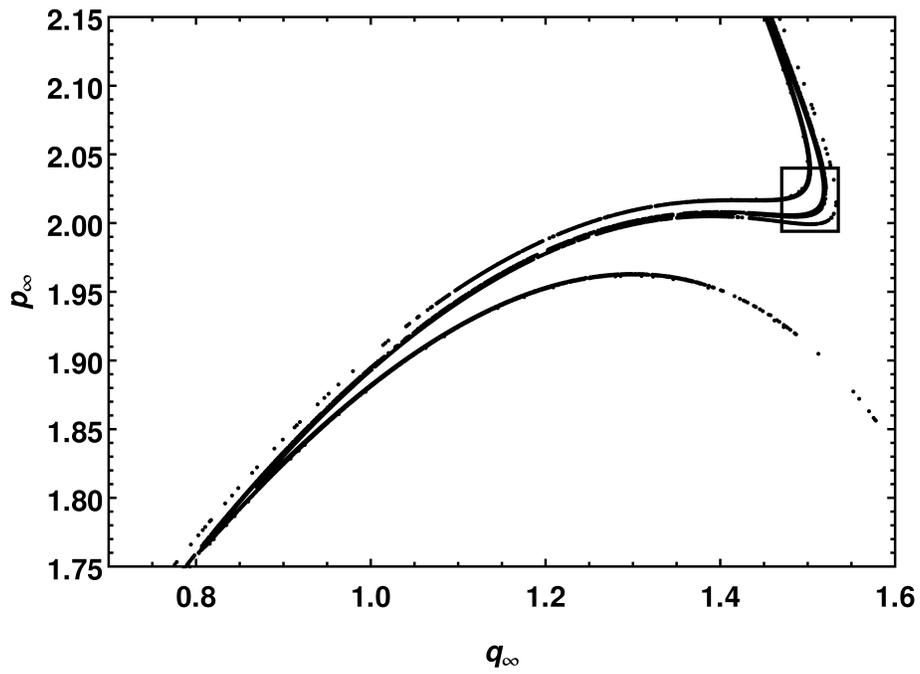}
  \caption{Limiting values of $q_{\infty},p_{\infty}$ for the map ${\cal {M}}_3$ when $\omega=1.2902$. 
 They appear to lie on a strange attractor.
}
\end{figure}

\begin{figure}[htp]
  \centering
  \includegraphics*[width=5.5in]{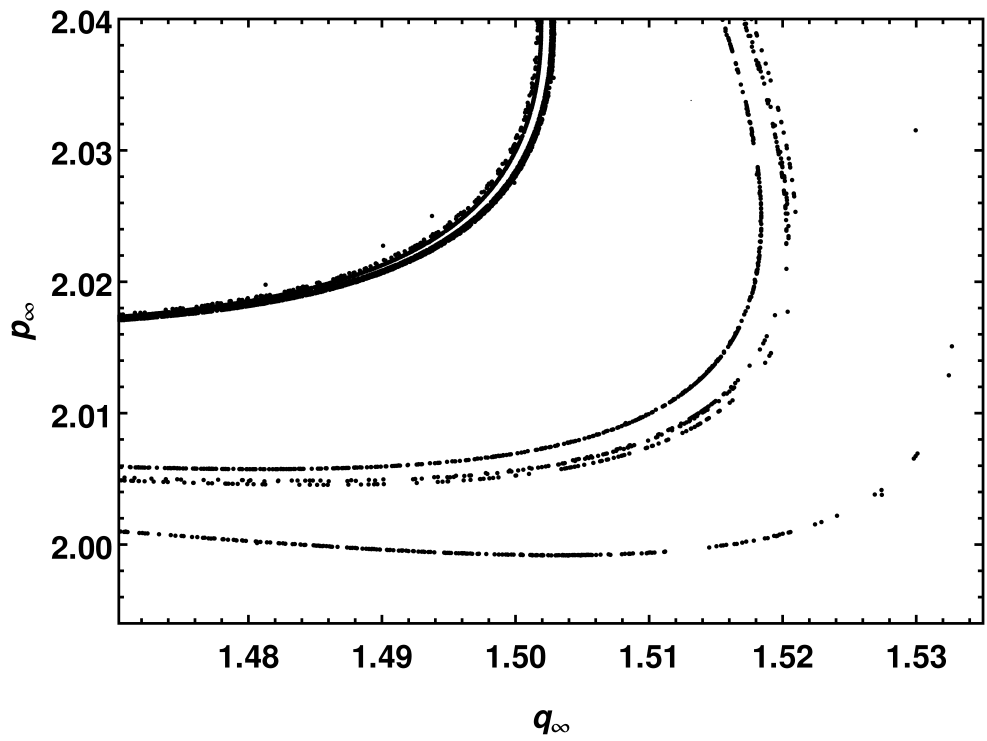}
  \caption{
Enlargement of boxed portion of Figure 22 illustrating the beginning of self-similar fractal structure.}
\end{figure}

\subsubsection{Performance of ${\cal{M}}_5$}

Figure 24 shows the ${\cal{M}}_5$ counterpart to Figure 15  
produced using ${\cal{M}}_8$.  Now there is improved quantitative agreement as well as qualitative agreement.  Also, there are now three features near $\omega=1.265$ that resemble those seen in Figures 12 and 15. 

Figures 25 and 26 show the  ${\cal{M}}_5$ counterparts to Figures 17 and 18 produced using ${\cal{M}}_8$. Again there is improved quantitative agreement.

\begin{figure}[htp]
  \centering
  \includegraphics*[width=4.5in]{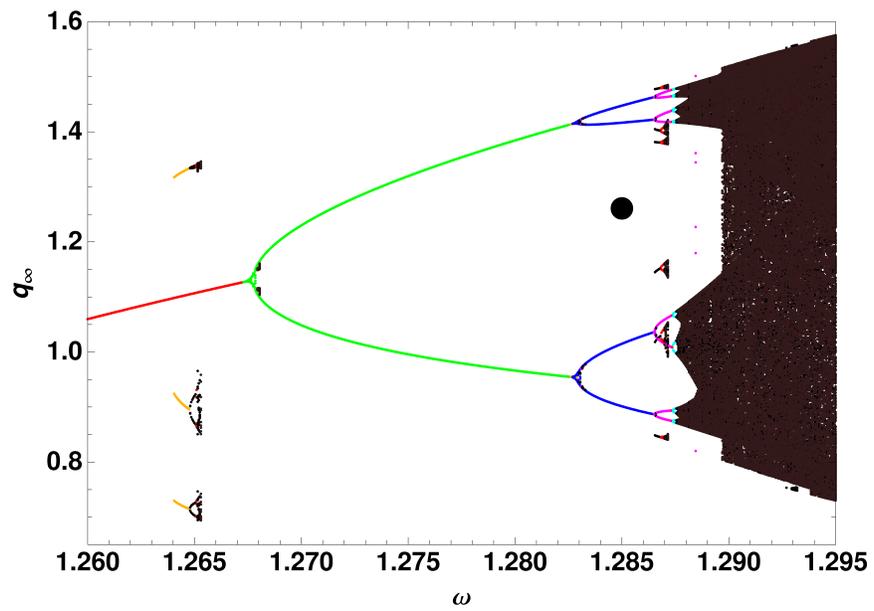}
  \caption{
Partial Feigenbaum diagram for the map  ${\cal {M}}_5$. The black dot marks the point about which ${\cal M}$ is expanded to yield ${\cal {M}}_5$}.
\end{figure}

\begin{figure}[htp]
  \centering
  \includegraphics*[height=4.5in]{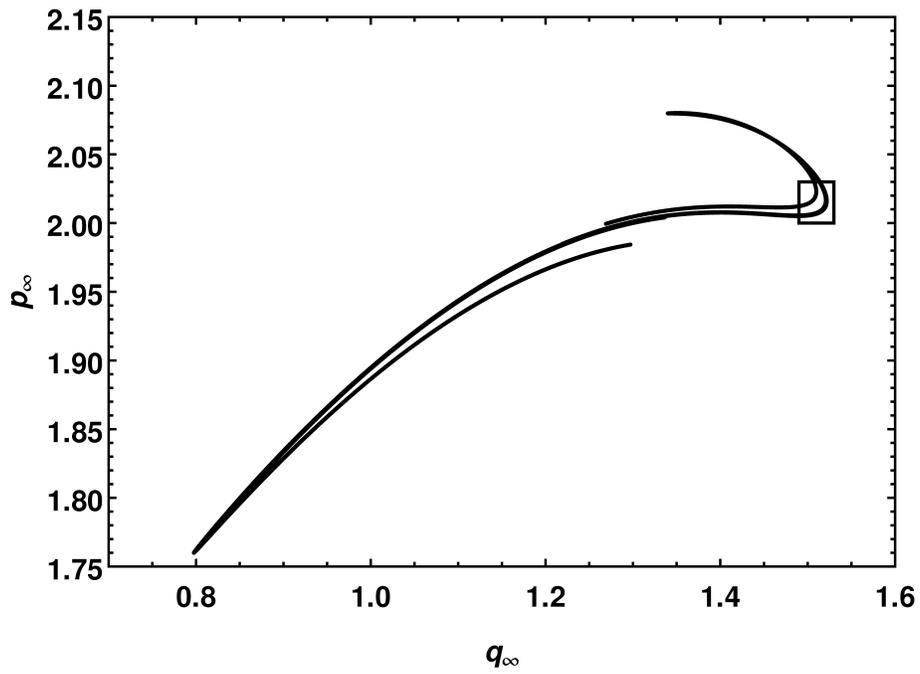}
  \caption{Limiting values of $q_{\infty},p_{\infty}$ for the map ${\cal {M}}_5$ when $\omega=1.2902$. They appear to lie on a   strange attractor.}
\end{figure}

\begin{figure}[htp]
  \centering
  \includegraphics*[width=5.5in]{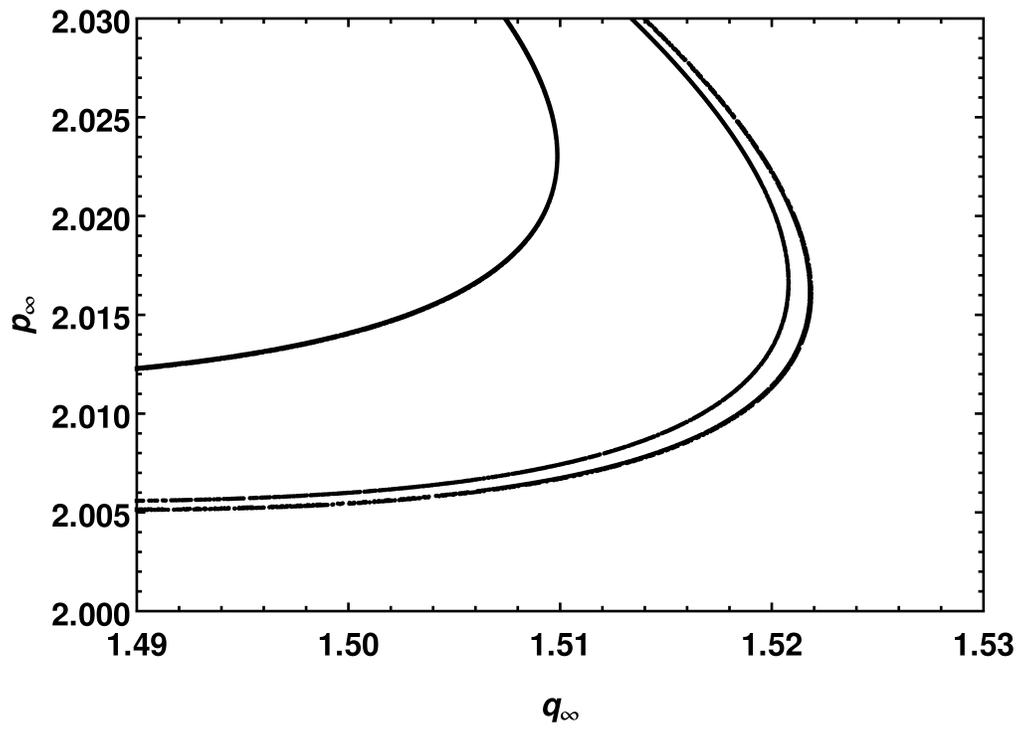}
  \caption{
Enlargement of boxed portion of Figure 25 illustrating the beginning of self-similar fractal structure.}
\end{figure}

\newpage

\section{Concluding Summary and Discussion} 
\setcounter{equation}{0}
Poincar\'{e} analyticity (and its generalization) implies that transfer maps $\cal M$ arising from ordinary differential equations can be expanded as Taylor series in the initial conditions and also in whatever parameters may be present.  Section 1 described the complete variational equations, and described how the determination of these expansions is equivalent to solving the complete variational equations. Section 2 provided an overview of the properties of the stroboscopic transfer map $\cal M$ for the Duffing equation.  Section 3 described  examples of how $n^{th}$ degree approximations ${\cal{M}}_n$ to  $\cal M$ (including parameter dependence) could reproduce various features of the exact $\cal M$.  In particular it illustrated, remarkably,  that ${\cal{M}}_8$ produced an infinite period doubling cascade and apparent strange attractor that closely resembled those of the exact map.  It also illustrated how the accuracy of ${\cal{M}}_n$ improves with increasing $n$.

We have seen that there are situations in which a truncated Taylor map well reproduces results obtained by the integration of differential equations.  This is comforting since the behavior of polynomial maps, because such maps can easily be evaluated repeatedly, is often studied in detail with the hope that the behavior of such maps is illustrative of what can be expected for maps in general, including the maps that arise from integrating differential equations.  

In view of this success, one might wonder if there are situations in which the use of truncated Taylor maps could replace or at least complement direct numerical integration.  There is, of course, the question of convergence for Taylor series, and the convergence domain is related to the (generally unknown) singularity structure of the solution to the differential equation in the complex domain [10].  However, if satisfactory approximation can be illustrated by the comparison of numerical integration results with truncated Taylor results for representative solutions in some domain, then the use of truncated Taylor maps to find additional results may be faster than continued numerical integration.  

For example, in the case of the Duffing equation, although the determination of the relevant $h_a^r(t)$ requires the simultaneous numerical integration of a large number of differential equations, these equations need be integrated over only one drive period.  Once the truncated Taylor series stroboscopic map has been found, its evaluation for any phase space point and any parameter value is essentially free.  All that is required is the evaluation of two $n$-degree polynomials (one for $\zeta^f_1$ and one for $\zeta^f_2$, the deviation variables associated with $q^f$ and $p^f$, respectively) in three variables ($\zeta_1^i$, $\zeta_2^i$, and $\zeta^i_3$).  By contrast, the direct construction of a Feigenbaum diagram requires the integration of the Duffing equation for a large number of drive periods and a large number of parameter values.  And, determination of the strange attractor associated with the Duffing equation requires the integration of the Duffing equation over thousands of drive periods. 

Suppose $T_2$ is the time required to integrate two equations over a drive period.  In our example, it is the time required to integrate the Duffing pair of differential equations (2.6) over one drive period.  Suppose $T_{N_e}$ is the time required to integrate $N_e$ equations over one drive period.  Let $L(m,n)$ be the number of monomials of degree $0$ through $n$ in $m$ variables.  It is given by the binomial coefficient
\begin{equation}
L(m,n)=\binom{m+n}{n}.
\end{equation} 
See [10].  When working with $m$ variables through terms of degree $n$, the number $N_e$ of differential equations to be integrated to determine the relevant functions $h^r_a(t)$ is given by the relation
\begin{equation}
N_e=mL(m,n),
\end{equation}
which amounts to 
\begin{equation}
N_e=3L(3,8)= 3\times165=495
\end{equation}
in the case of  ${\cal {M}}_8$ for the Duffing equation including parameter dependence.  We have found in our numerical studies that there is the approximate scaling relation
\begin{equation}
T_{N_e}\simeq(N_e/2)T_2
\end{equation}
for $n\le9$.  That is, the computation time scales with the number of equations to be integrated.  We conclude that in this example the use of ${\cal {M}}_8$ becomes advantageous once the number of drive periods times the number of parameter values exceeds $495/2\simeq250$.

With regard to providing complementary information, it is common practice to integrate the first degree variational equations in order to establish the {\em linear} stability of solutions.  Integration of the higher degree variational equations, including possible parameter dependence,  provides information about {\em nonlinear} behavior/stability.    As examples, such information is required for the control of orbits in accelerators and the understanding and control of aberrations in optical systems.

In conclusion, there are applications for which use of the higher degree variational equations is advantageous, and the whole subject of the usefulness of truncated Taylor maps merits continued study.

\section{Acknowledgments} 
D.~Kaltchev wishes to thank his colleagues from TRIUMF and CERN, especially Richard Abram Baartman, for their interest and support.


\begin{thebibliography}{99}
\bibitem{} J. Barrow-Green, {\em Poincar\'{e} and the Three Body Problem}, American Mathematical Society (1997).
\bibitem{} F. Browder, Edit., {\em The Mathematical Heritage of Henri Poincar\'{e}, Proceedings of Symposia in Pure Mathematics of the American Mathematical Society} {\bf{39}}, Parts 1 and 2,  American Mathematical Society (1983).
\bibitem{} H. Poincar\'{e}, {\em New Methods of Celestial Mechanics,} Parts 1, 2, and 3.  (Originally published as  Les M\'{e}thodes nouvelles de la M\'{e}chanique c\'{e}leste.)  American Institute of Physics {\em History of Modern Physics and Astronomy}, Volume 13, D. L. Goroff, Edit., American Institute of Physics (1993).  See page 145 of the Editor's Introduction for a statement of Poincar\'{e}'s holomorphic lemma, and Sections \S20 through \S27 for Poincar\'{e}'s proof.
\bibitem{} Francis J. Murray and Kenneth S. Miller, {\em Existence Theorems for Ordinary Differential Equations}, New York University Press and Interscience Publishing Co. (1954).
\bibitem{} Y. Ilyashenko and S. Yakovenko, {\em Lectures on Analytic Differential Equations}, American Mathematical Society (2008).  See Theorem 1.1.
\bibitem{} E. Coddington and N. Levinson, {\em Theory of Ordinary Differential Equations}, McGraw-Hill (1955).  See Chapter 1, Section \S 8, Theorem 8.4.
\bibitem{} W. Walter, {\em Ordinary Differential Equations}, Springer (1998).  See Section \S13.
\bibitem{} J. Guckenheimer and P. Holmes, {\em Nonlinear Oscillations, Dynamical Systems, and Bifurcations of Vector Fields}, Springer (1983).
\bibitem{} D. Kaltchev and A. J. Dragt, ``Poincar\'{e} Analyticity and the Complete Variational Equations", {\em Physica D: Nonlinear Phenomena} $\bf{242}$ (2013).
\bibitem{} A. J. Dragt, {\em Lie Methods for Nonlinear Dynamics with Applications to Accelerator Physics},  (2012), available at http://www.physics.umd.edu/dsat/

\end{thebibliography}
\end{document}